\documentclass[preprint,aps,amssymb,showpacs,superscriptaddress,nofootinbib]{revtex4}
\usepackage{graphicx}
\newcommand{\del}{\partial}

\begin{document}

\title{Gravity Asymptotics with Topological Parameters}

\author{Sandipan Sengupta}
\email{sandipan@rri.res.in}
\affiliation{Raman Research Institute\\
Bangalore-560080, INDIA.}

\begin{abstract}
In four dimensional gravity theory, the Barbero-Immirzi parameter has a topological origin, and can be identified as the coefficient multiplying the Nieh-Yan topological density in the gravity Lagrangian, as proposed by Date et al.\cite{date}. Based on this fact, a first order action formulation for spacetimes with boundaries is introduced. The bulk Lagrangian, containing the Nieh-Yan density, needs to be supplemented with suitable boundary terms so that it leads to a well-defined variational principle. Within this general framework, we analyse spacetimes with and without a cosmological constant.

For locally Anti de Sitter (or de Sitter) asymptotia, the action principle has non-trivial implications. It admits an extremum for all such solutions provided the SO(3,1) Pontryagin and Euler topological densities are added to it with fixed coefficients. The resulting Lagrangian, while containing all three topological densities, has  only one independent topological coupling constant, namely, the Barbero-Immirzi parameter. In the final analysis, it emerges as a coefficient of the SO(3,2) (or SO(4,1)) Pontryagin density, and is present in the action only for manifolds for which the corresponding topological index is non-zero.
 
\end{abstract}


\maketitle

%%%%%%%%%%%%%%%%%%%%%%%%%%%%%%%%%%%%%%%%%%%%%%%%%%%%%%%%%%%%%%%%%%%%%%%
\section{Introduction}
In four spacetime dimensions, the theory of gravity has three independent topological parameters\cite{kaul}. In the first order action formulation, these are associated with three topolological densities, namely, the Nieh-Yan, Euler and Pontryagin\cite{kaul,nieh,hehl}. Among these, the Nieh-Yan density\cite{nieh} shows up only in first order gravity where the tetrad and spin-connection are treated as independent variables. In terms of these basic fields, the Nieh-Yan density $I_{NY}$ is locally defined as:
\begin{eqnarray}\label{NY}
  I_{NY}&~=~&\epsilon^{\mu\nu\alpha\beta}\left[(D_{\mu}(\omega)e_{\nu}^I) ~(D_{\alpha}(\omega)e_{\beta I})~-~
\frac{1}{2}e_{\mu}^I e_{\nu}^J R_{\alpha\beta IJ}(\omega)\right]\nonumber\\
&~=~&\del_{\mu}\left[\epsilon^{\mu\nu\alpha\beta}e^{I}_{\nu}\left(D_{\alpha}(\omega)e_{\beta I}\right)\right]
  %&~=~&\del_{\mu}[\epsilon^{\mu\nu\alpha\beta}e^{I}_{\nu}(\del_{\alpha}e_{I\beta}+\omega_{\alpha IJ}e_{\beta}^{J})],
\end{eqnarray}
where, we define the covariant derivative $D_\mu(\omega)$ as: $D_\mu(\omega) e_\nu^I~=~\del_\mu e_\nu^I~+~\omega_{\mu}^{~IJ}e_{\nu J}$~. 
%In the usual metric (or second order) formulation with the symmetric %Christoffel symbol, the Nieh-Yan density is identically zero. 
It has been noted that this topological class typically appears in the context of canonical SU(2) formulation of gravity\cite{barbero,complex} with fermionic matter, for which the corresponding effective Lagrangians, also known as the generalised Holst Lagrangians, contain this term\cite{mercuri,kaul1}. However, the real importance of the Nieh-Yan density in the canonical theory of gravity with or without matter was first elucidated by Date et al.\cite{date}, who developed a Hamiltonian formulation of gravity based on a Lagrangian (density) made up of the Hilbert-Palatini and Nieh-Yan terms:
\begin{eqnarray}\label{L-NY}
  L(e,\omega)~=~\frac{1}{2\kappa}e\Sigma^{\mu\nu}_{IJ}R_{\mu\nu}^{~~IJ}(\omega)~+~\eta I_{NY}
\end{eqnarray} 
Here, 
$\Sigma_{IJ}^{\mu\nu} ~ = ~
\frac{1}{2}~(e_{I}^{\mu}e_{J}^{\nu}-e_{J}^{\mu}e_{I}^{\nu}) ,~ R^{~~~
  IJ}_{\mu\nu}(\omega) ~ = ~ \del_{[\mu} \omega_{\nu]}^{~IJ} +
\omega_{[\mu}^{~IK}\omega_{\nu]K}^{~ ~ ~J},~e~=~det(e_\mu^I)$ and $\kappa$ is the Gravitational constant. 
In the second term, the constant real coefficient $\eta$ is known as the (inverse of) Barbero-Immirzi parameter.
The resulting theory, while leading to a real SU(2) gauge theory of gravity exactly as in the earlier formulation of Holst\cite{holst,sa}, allows the introduction of any arbitrary matter-coupling without requiring any further modifications in the Lagrangian\cite{date,sengupta}. This is so because the Nieh-Yan topological density, being a total divergence, does not affect the equations of motion of Hilbert-Palatini gravity whether or not matter is coupled to the theory. In addition, the Lagrangian (\ref{L-NY}) provides a clear topological interpretation for the Barbero-Immirzi parameter $\eta$ which multiplies the Nieh-Yan density (for subsequent discussions on the topological origin of $\eta$ based on this fact, see \cite{sengupta1,mercuri1}). This is in contrast to the Holst formulation where $\eta$ appears as a coefficient of the Holst term in the Lagrangian\cite{holst}:
\begin{eqnarray}\label{L-holst}
  L(e,\omega)~=~\frac{1}{2\kappa}e\Sigma^{\mu\nu}_{IJ}R_{\mu\nu}^{IJ}(\omega)~+~\frac{\eta}{2}
  e\epsilon^{IJKL} \Sigma^{\mu\nu}_{IJ}R_{\mu\nu KL}(\omega)
\end{eqnarray}
Since the Holst term is not a topological density, it does not elucidate the topological origin of $\eta$. Although there are some special instances where the Holst term captures the same topological information as the Nieh-Yan density, the formulation with Nieh-Yan is more general, even in pure gravity. This is because a vanishing Nieh-Yan density necessarily implies a vanishing Holst density, although the converse is not true. In fact, the torsional configurations in pure gravity as studied by Chandia and Zanelli in \cite{zanelli1} constitute an example where the corresponding Nieh-Yan topological index is non-trivial, even though the Holst density vanishes. In the presence of matter-coupling, the Holst term needs matter-dependent modifications which are not universal\cite{mercuri,kaul1}. Thus, the action principle based on (\ref{L-NY}) supercedes the framework of Holst\cite{date,kaul,sengupta}.

However, the analysis in \cite{date} is relevant for manifolds which are either compact without boundaries or have boundaries where the surface terms do not contribute. For spacetimes with general boundaries, where the surface terms are really relevant to the analysis, the significance of the Nieh-Yan density is yet to be understood. Such an exercise is important from the perspective discussed above, which suggests that all matter couplings in gravity theory should be treated in a universal manner, and the Barbero-Immirzi parameter should have a direct topological interpretation, even within the action formulation for spacetimes with non-trivial asymptotia.

In the first order gravity asymptotics\cite{ashtekar}, the role of the Barbero-Immirzi parameter has been a topic of active interest for a while\cite{corichi}. However, all these earlier works are based on the Holst formulation.
The suggestion that $\eta$ might show up through the Nieh-Yan density in the action principle for manifolds with boundaries was recently made in ref.\cite{norbert}. Their analysis, which also is based on the Holst framework and deals with closed boundaries, proposes a surface term involving the `torsional Chern-Simons density'\footnote{The Nieh-Yan topological density can be written as a total divergence, as in eqn.(\ref{NY}): $I_{NY}=\del_\mu J^\mu_{NY}$. We define $J^{\mu}_{NY}=\epsilon^{\mu\nu\alpha\beta}e^{I}_{\nu}\left(D_{\alpha}(\omega)e_{\beta I}\right)$ as the `torsional Chern-Simons density'.}. The full Lagrangian in this framework contains both the Holst and Nieh-Yan densities.

Here, in this short note, we set up an action principle based on a Lagrangian containing the Hilbert-Palatini and Nieh-Yan densities for manifolds with boundaries, and study the implications. First, we analyse the case for Dirichlet boundaries. A well-known example of such geometries is the asymptotically flat spacetime. Next, we consider spacetimes which has asymptotic boundaries with constant negative (or positive) curvature (locally). When the asymptotic boundary is the only boundary, these are known as asymptotically locally Anti de Sitter (ALADS) geometries\cite{zanelli}. Although the analysis of boundary terms for this class of asymptotia has a long history, the fact that topological densities play an important role in the corresponding action formulation was first demonstrated by Aros et al.\cite{zanelli}. They showed that the boundary term corresponding to the Hilbert-Palatini density for such asymptotia can be written as the Euler topological density, multiplied by a coefficient fixed in terms of the gravitational and cosmological constants. Thus, this theory has no independent topological parameter. Here, we find that with the inclusion of the Nieh-Yan density, the Lagrangian admits an extremum for all ALADS solutions provided the Pontryagin topological density is included in it with a fixed coefficient, i.e. the Barbero-Immirzi parameter $\eta$. Thus, our analysis demonstrates that the most general action principle for such asymptotic geometries has $\eta$ as the only topological parameter, while containing all three independent topological densities which exist in four-dimensional gravity theory, namely, the Nieh-Yan, Pontryagin and Euler. It is also important to note that our analysis does not require the introduction of the Holst term in the Lagrangian, unlike the earlier formulations\cite{corichi,norbert}.

In the next section, we introduce the action principle containing the Nieh-Yan density and apply it to spacetimes with Dirichlet boundaries. Next, we extend this analysis for locally Anti de Sitter (or de Sitter) asymptotic boundaries and study the consequences. The last section contains a few relevant remarks.
\vspace{.4cm}

\section{Action principle }
\vspace{.2cm}
For a four-dimensional spacetime manifold $M$ whose boundary is $\del M$, we propose the following Lagrangian density for pure gravity:
\begin{equation} \label{L}
  L(e,\omega)~=~\frac{1}{2\kappa}e\Sigma^{\mu\nu}_{IJ}R_{\mu\nu}^{~~IJ}~+~\eta I_{NY}~+~B
\end{equation} 
where, the Nieh-Yan density $I_{NY}$ is defined as in (\ref{NY}), and $B$ is a surface term, depending on the fields at the boundary. $B$ can be fixed by
demanding a well-defined variational principle for the action\footnote{Issues related to the convergence of the action are not discussed here. For relevant discussions in this regard, see \cite{corichi,zanelli,aros}.}. The above action principle can be  generalised for any arbitrary matter-coupling in a straightforward manner, by simply adding the matter Lagrangian as it is (along with the corresponding boundary term), without changing the Nieh-Yan density. 

Variation of (\ref{L}) with respect to the independent fields $e_\mu^I$ and $\omega_\mu^{IJ}$ leads to:
%\newpage
\begin{eqnarray}\label{L-B}
\delta L(e,\omega)~&=&~\frac{1}{4\kappa} \epsilon^{\mu\nu\alpha\beta}\epsilon_{IJKL}
\left[e_\mu^I R_{\alpha\beta}^{~~KL}\delta e_\nu^J~+~2e_\mu^I (D_\alpha e_\beta^J) 
\delta\omega_{\nu}^{KL}\right]~+~\delta B  \nonumber\\
&~+&~\del_{\mu}\left[\epsilon^{\mu\nu\alpha\beta}~\left(\frac{1}{4\kappa}\epsilon_{IJKL} 
e_\alpha^I e_\beta^J \delta\omega_{\nu}^{KL}~+~\eta~(e_\nu^I e_\beta^J \delta \omega_{\alpha IJ}~+~
2(D_{\alpha}e_{\beta I}) \delta e_\nu^I)\right)\right]
\end{eqnarray}
where, we have used the identity: $e\Sigma^{\mu\nu}_{IJ}=\frac{1}{4}\epsilon^{\mu\nu\alpha\beta}\epsilon_{IJKL} \Sigma_{\alpha\beta}^{KL}~$.
While the first two terms in the parenthesis correspond to the equations of motion, the remaining
ones contribute at the boundary of the spacetime. 
It follows from (\ref{L-B}) that the Lagrangian density (\ref{L}) will have an extremum for all solutions subject to suitable boundary conditions if the following holds:
\begin{eqnarray}\label{deltaB}
\delta B~=~-\epsilon^{abc}\left(\frac{1}{4\kappa}\epsilon_{IJKL} e_{a}^{K}e_{b}^{L}
~-~\eta e_{a I}e_{b J}\right)\delta \omega_{c}^{IJ}
\end{eqnarray}
where, the indices a,b,c etc. correspond to the coordinates on the three dimensional boundary manifold $\del M$ and $\epsilon^{abc}$ is the Levi-Civita tensor density induced at the boundary.
In what follows next, we find out the explicit form of the boundary term $B$ for spacetimes which have Dirichlet and locally ADS asymptotia, respectively. If the spacetime has boundaries other than the asymptotic one, a Dirichlet condition on the spin-connection would be assumed on such boundaries\cite{aros}: 
\begin{eqnarray}
\delta \omega_{a'}^{IJ}~=~0,
\end{eqnarray}
with $a'$ denoting the indices corresponding to the coordinates on the non-asymptotic three-boundary. 
As is evident from (\ref{deltaB}), for this boundary condition, $\delta B$
vanishes. Thus, the non-asymptotic surfaces do not contribute to the boundary term $B$. Hence, it is enough to consider only the asymptotic boundary, as would be done in the rest of the paper.

At this point, it is important to emphasize that the boundary contribution in (\ref{deltaB}) corresponding to the Nieh-Yan density 
is exactly the same as the one for the Holst term\cite{corichi}, as appearing within the Holst action formulation based on the Lagrangian density (\ref{L-holst}).
This fact can be explicitly checked by taking a variation of the Holst density $L_H$:
\begin{eqnarray*}
\delta L_H~&=&~~\frac{1}{2}\delta \left(e \epsilon^{IJKL} e^\mu_I e^\nu_J R_{\mu\nu KL} \right)\\
~&=&-\frac{1}{2}\delta \left(\epsilon^{\mu\nu\alpha\beta}e_\mu^I e_\nu^J R_{\alpha\beta IJ} \right)\\
~&=&-\frac{1}{2}\epsilon^{\mu\nu\alpha\beta} \left[e_\nu^J R_{\alpha\beta IJ}\delta e_\mu^I~-~2 e_\nu^J (D_{\alpha}
e_\mu^I) \delta \omega_{\beta IJ}  \right]~-~ \del_{\alpha}
\left(\epsilon^{\mu\nu\alpha\beta} e_\mu^I e_\nu^J \delta \omega_{\beta IJ} \right)
\end{eqnarray*}
Here, in the second line we have used the identity: $e\epsilon^{IJKL} e^\mu_I e^\nu_J~=- \epsilon^{\mu\nu\alpha\beta}e_\alpha^K e_\beta^L~ $.
While the first two terms above contribute to the equations of motion (these contributions do not affect the Hilbert-Palatini equations of motion), the total divergence term contains the boundary contribution $\delta B_H$ corresponding to the Holst term:
\begin{eqnarray}\label{delta-holst}
\delta B_H~=~-\epsilon^{abc} e_{aI} e_{bJ} \delta \omega_{c}^{IJ}
\end{eqnarray}
Comparing (\ref{delta-holst}) with the second-term in (\ref{deltaB}), we conclude that the boundary contributions $B_{NY}$ and $B_H$ corresponding to the Nieh-Yan and Holst densities, respectively, are exactly the same for pure gravity upto a sign\footnote{For matter-coupling leading to a non-vanishing torsion, e.g.fermions, these two differ by $B_{NY}-B_H \approx \epsilon^{abc} D_a e_b^I \delta e_{cI}$ (see eqn. (\ref{L-B})), which is non-vanishing for boundaries for which $\delta e_a^I \neq 0$.}. Thus, in order to introduce the Barbero-Immirzi parameter $\eta$ in the theory, the inclusion of Nieh-Yan density with the coefficient $\eta$ is sufficient. One does not need a (further) addition of the Holst density to the Lagrangian (\ref{L}). Also, as we will see later, the boundary term $B_{NY}$ corresponding to the Nieh-Yan is gauge invariant as it is, when appropriate boundary conditions are used. Thus, although one can still work within an action principle containing both the Nieh-Yan and Holst terms multiplied by the same coefficient $\eta$ as in ref.\cite{norbert}, this is by no means necessary. Such an approach, however, obscures the topological interpretation of $\eta$, and should be avoided from our viewpoint. 

 \subsection{Dirichlet boundary}
At the boundary at infinity, we assume a Dirichlet condition on the tangential components of the tetrad, keeping $\delta \omega_{\mu}^{IJ}$ arbitrary: 
\begin{eqnarray}\label{bc}
\delta e_a^I~=~0
\end{eqnarray}
Note that asymptotically flat spacetimes constitute an example of such geometries\cite{ashtekar}.

From (\ref{deltaB}), it follows that for this boundary condition (\ref{bc}), the surface term B can be written as:
\begin{eqnarray}\label{B}
B=~-\epsilon^{abc}\left(\frac{1}{4\kappa}\epsilon_{IJKL} e_{a}^{K}e_{b}^{L}
~-~\eta e_{aI}e_{bJ}\right)\omega_{c}^{IJ}
\end{eqnarray}
In the above, the first and second terms correspond to the boundary contributions from the 
Hilbert-Palatini and Nieh-Yan densities, respectively. 
%This is in contrast to the action formulation discussed in (\cite{norbert}).

Notice that according to the boundary condition (\ref{bc}), the tangential components of the tetrad are fixed at the boundary. This implies that in the asymptotic region, the only consistent (infinitesimal) gauge transformations are those which are trivial. Thus, the boundary term $B$ is gauge-invariant. 

\subsection{Locally AdS (dS) boundary}
Next, we consider a spacetime with an asymptotic boundary which locally has constant negative or positive curvature. Assuming that this is the only boundary, we call this asymptotically locally Anti de Sitter (ALADS) or de Sitter spacetime borrowing the standard terminology\cite{zanelli}. For such geometries, we set up an action principle containing 
the Nieh-Yan term in the Lagrangian density. Note that the asymptotic boundary condition used here is not equivalent to the Dirichlet condition (\ref{bc}) as used in the earlier case. 

For ALADS spacetimes, the curvature tensor at the boundary at infinity locally satisfies the following relation\cite{zanelli,aros}:
\begin{eqnarray}\label{R-ads}
R_{ab}^{~~IJ}~+~\frac{1}{l^2}e_{[a}^{I}e_{b]}^{J}~=~0
\end{eqnarray}
where, the AdS radius $l$ is related to the cosmological constant $\Lambda$ as: $\Lambda=-\frac{3}{l^2}$. Although we present the explicit computations below for Anti de Sitter asymptotic boundaries, taking $\Lambda$ to be negative, our analysis also applies to the de Sitter case, which corresponds to a positive $\Lambda$.
 
For pure gravity with a negative cosmological constant, the Lagrangian density for four-dimensional manifolds with a boundary is given by:
 \begin{eqnarray}
 L(e,\omega)~=~\frac{1}{8\kappa} \epsilon^{\mu\nu\alpha\beta}\epsilon_{IJKL}
\left(e_\mu^Ie_\nu^J R_{\alpha\beta}^{~~KL}~+~\frac{1}{l^2}e_\mu^Ie_\nu^J e_\alpha^K e_\beta^L\right)~+~B
 \end{eqnarray}
where, $B$ is a functional of the fields at the boundary.
According to the general proposal presented earlier, we introduce the Barbero-Immirzi parameter as a topological coupling constant in this theory through the Nieh-Yan density:
 \begin{eqnarray}\label{L-ads}
 L(e,\omega)~=~\frac{1}{8\kappa} \epsilon^{\mu\nu\alpha\beta}\epsilon_{IJKL}
\left(e_\mu^Ie_\nu^J R_{\alpha\beta}^{~~KL}~+~\frac{1}{l^2}e_\mu^Ie_\nu^J e_\alpha^K e_\beta^L\right)~+~\eta I_{NY}~+~B
\end{eqnarray}
where, $I_{NY}$ is defined in (\ref{NY}).
Varying the Lagrangian density above with respect to the independent fields $e_\mu^I$ and $\omega_{\mu}^{IJ}$, we obtain: 
\begin{eqnarray}\label{deltaL}
\delta L(e,\omega)~&=&~\frac{1}{4\kappa} \epsilon^{\mu\nu\alpha\beta}\epsilon_{IJKL}
\left[e_\mu^I \left(R_{\alpha\beta}^{~~KL}+\frac{1}{l^2}e_{[\alpha}^K e_{\beta]}^L\right)\delta e_\nu^J~+~2e_\mu^I (D_\alpha e_\beta^J) 
\delta\omega_{\nu}^{KL}\right]\nonumber\\
&~+&~\del_{\mu}\left[\epsilon^{\mu\nu\alpha\beta}\left(\frac{1}{4\kappa}\epsilon_{IJKL} 
e_\alpha^I e_\beta^J \delta\omega_{\nu}^{KL}~+~\eta\left(e_{\nu I} e_{\beta J} \delta \omega_{\alpha}^{IJ}~+~
2(D_{\alpha}e_{\beta I}) \delta e_\nu^I\right)\right)\right]\nonumber\\
&~+&~\delta B
\end{eqnarray}
The first line above corresponds to the equations of motion, while the rest contain the boundary terms. The Lagrangian density (\ref{L-ads}) admits a well-defined variational principle, provided the total contribution at the boundary vanishes. This implies:
\begin{eqnarray}\label{deltaB-ads}
\delta B~=~-\epsilon^{abc}\left(\frac{1}{4\kappa}\epsilon_{IJKL} e_{a}^{K}e_{b}^{L}
~-~\eta e_{aI}e_{bJ}\right)\delta \omega_{c}^{IJ}~+~2\eta (D_{\alpha}e_{\beta I}) \delta e_\nu^I
\end{eqnarray}
Using the equations of the motion (at the ALADS boundary), which are given by eqn.(\ref{R-ads}) and the vanishing of torsion, (\ref{deltaB-ads}) can be rewritten as:
\begin{eqnarray}\label{deltaB-final}
\delta B~=~-\frac{l^2}{2}\epsilon^{abc}\left(\frac{1}{4\kappa}\epsilon_{IJKL} R_{ab}^{~~KL}\delta \omega_{c}^{IJ}
~-~\eta ~ R_{abIJ}\delta \omega_{c}^{IJ}\right)
\end{eqnarray}
Now notice that the two terms above are precisely the variations of the Chern-Simons densities $C_E=\frac{1}{2}\epsilon^{abc}\epsilon_{IJKL}\omega_a^{IJ}\left(\del_b\omega_c^{KL} ~+~\frac{2}{3}\omega_b^{KM}\omega_{cM}^{~~L}\right)$ and $C_P=\epsilon^{abc}\omega_{aIJ}\left(\del_b\omega_c^{IJ} ~+~\frac{2}{3}\omega_b^{IK}\omega_{cK}^{~~J}\right)$, corresponding to the Euler and Pontryagin terms, respectively: 
\begin{eqnarray*}
\delta C_E&~=&~\delta~\left[\frac{1}{2}\epsilon^{abc}\epsilon_{IJKL}\omega_a^{IJ}\left(\del_b\omega_c^{KL} ~+~\frac{2}{3}\omega_b^{KM}\omega_{cM}^{~~L}\right)\right]\\
&~=&~\frac{1}{2}\epsilon^{abc}\epsilon_{IJKL}R_{ab}^{IJ}\delta \omega_{c}^{KL}~~\\
\delta C_P&~=&~\delta~\left[\epsilon^{abc}\omega_{aIJ}\left(\del_b\omega_c^{IJ} ~+~\frac{2}{3}\omega_b^{IK}\omega_{cK}^{~~J}\right)\right]\\
&~=&~\epsilon^{abc}R_{abIJ}\delta \omega_{c}^{IJ}~~
\end{eqnarray*}
These come with fixed coefficients in (\ref{deltaB-final}), being completely determined in terms of $\kappa, l$ and $\eta$. 
Thus, the boundary contribution B in (\ref{L-ads}) can be written as: 
\begin{eqnarray}
B~=~-\frac{l^2}{2}\left(\frac{1}{2\kappa} C_E~-~\eta C_P\right)
\end{eqnarray}

To study the effect of gauge transformations on these boundary terms, we note that under 
a typical infinitesimal transformation of the form: $\delta_G \omega_{\mu}^{IJ}=D_\mu(\omega)\theta^{IJ}$, 
$C_E$ transforms as:
\begin{eqnarray*}
\delta_G C_E&~=&~\frac{1}{2}\epsilon^{abc}\epsilon_{IJKL}R_{ab}^{~~IJ}D_a(\omega)\theta^{KL}\\
&~=&~\frac{1}{2}\del_a\left[\epsilon^{abc}\epsilon_{IJKL}R_{ab}^{~~  IJ}\theta^{KL}\right]\\
&~=&~0
\end{eqnarray*}
where, in the second line we have used the Bianchi identity and in the third line we have assumed that the two-dimensional boundary of $\del M$ is such that the boundary contribution there vanishes. The gauge invariance of $C_P$ under infinitesimal transformations can be checked similarly.
%Notice that in order to determine the boundary term B, one does not need to assume a Dirichlet %condition on the boundary variations $\delta e_a^I$ as in the earlier section.

Now, addition of the Chern-Simons densities $C_E$ and $C_P$ at the boundary is equivalent to the addition of the Euler and Pontryagin topological densities $I_E$ and $I_P$ in the bulk theory. This can be demonstrated using the following identities:
\begin{eqnarray*}
I_E&~=&~\frac{1}{8}\epsilon^{\mu\nu\alpha\beta}\epsilon_{IJKL}R_{\mu\nu}^{~~IJ}R_{\alpha\beta}^{~~KL}~=~\frac{1}{2}\del_\mu \left[\epsilon^{\mu\nu\alpha\beta}\epsilon_{IJKL}\omega_\nu^{~IJ}\left(\del_\alpha\omega_\beta^{~KL} ~+~\frac{2}{3}\omega_\alpha^{~KM}\omega_{\beta M}^{~~~L}\right)\right]\\
I_P&~=&~\frac{1}{4}\epsilon^{\mu\nu\alpha\beta}R_{\mu\nu}^{~~IJ}R_{\alpha\beta IJ }~=~\del_\mu \left[\epsilon^{\mu\nu\alpha\beta}\omega_\nu^{~IJ}
\left(\del_\alpha\omega_{\beta IJ} ~+~\frac{2}{3}\omega_{\alpha I}^{~~K}\omega_{\beta KJ}\right)\right]
\end{eqnarray*}
Using these, the Lagrangian density (\ref{L-ads}) finally can be written as:
\begin{eqnarray}\label{L-final}
L(e,\omega)&~=&~\frac{1}{8\kappa} \epsilon^{\mu\nu\alpha\beta}\epsilon_{IJKL}
\left(e_\mu^Ie_\nu^J R_{\alpha\beta}^{~~KL}~+~\frac{1}{l^2}e_\mu^Ie_\nu^J e_\alpha^K e_\beta^L\right)~+~\frac{l^2}{4\kappa}I_E~+~\eta I_{NY}~-~\frac{\eta l^2}{2} I_P\nonumber\\
&~=&~\frac{1}{8\kappa} \epsilon^{\mu\nu\alpha\beta}\epsilon_{IJKL}
\left(e_\mu^Ie_\nu^J R_{\alpha\beta}^{~~KL}~+~\frac{1}{l^2}e_\mu^Ie_\nu^J e_\alpha^K e_\beta^L\right)~+~\frac{l^2}{32\kappa}
\epsilon^{\mu\nu\alpha\beta}\epsilon_{IJKL}R_{\mu\nu}^{~~IJ}R_{\alpha\beta}^{~~KL}\nonumber\\
&~+&~\eta \epsilon^{\mu\nu\alpha\beta}\left((D_{\mu} e_\nu^I)~ (D_\alpha e_{\beta I})
~-~\frac{1}{2}e_\mu^I e_\nu^J R_{\alpha\beta IJ} \right)
~-~\frac{\eta l^2}{8}\epsilon^{\mu\nu\alpha\beta}R_{\mu\nu}^{~~IJ}R_{\alpha\beta IJ}
\end{eqnarray}
By construction, this action principle has an extremum for all ALADS solutions. The striking fact is that all the three topological densities which exist in four dimensional gravity theory appear in the final Lagrangian density. However, not all three coefficients are independent. While the Euler coefficient is completely fixed in terms of $\kappa$ and $\Lambda$, the Nieh-Yan and Pontryagin densities both appear with the coefficient $\eta$. Thus, the Barbero-Immirzi parameter $\eta$ emerges as the only independent topological coupling constant in this theory.
\vspace{.4cm}

{\bf Emergence of SO(3,2) Pontryagin density:}
\vspace{.3cm}

Let us observe that in eqn.(\ref{L-final}), the Nieh-Yan and Pontryagin densities come with weights such that they can be combined into a single topological density, namely the SO(3,2) Pontryagin (for $\Lambda>0$, the corresponding gauge group becomes SO(4,1)). This can be understood with the help of the following construction\cite{zanelli1}. First, we define the SO(3,2) spin connection $W_\mu^{AB}$ built out of the tetrad $e_\mu^I$ and the SO(3,1) spin connection $\omega_\mu^{IJ}$, where the SO(3,2) indices A,B,.. run from 0 to 4 and the SO(3,1) indices I,J,.. run from 0 to 3:
\begin{eqnarray*}
W_\mu^{IJ}~=~\omega_\mu^{IJ};~~W_\mu^{I4}~=~\frac{1}{l}e_\mu^{I}
\end{eqnarray*}
The components of the SO(3,2) field-strength $F_{\mu\nu}^{AB}(W)=\del_{[\mu}W_{\nu]}^{AB}+W_{[\mu}^{AC}W_{\nu]C}^{~~B}$ thus become:
\begin{eqnarray*}
F_{\mu\nu}^{~~IJ}(W)~=~R_{\mu\nu}^{~~IJ}(\omega)~+~\frac{1}{l^2}e_{[\mu}^{I}e_{\nu]}^{J};~~F_{\mu\nu}^{~~4I}(W)~=~\frac{1}{l}D_{[\mu}(\omega)e_{\nu]}^I~=~\frac{2}{l}T_{\mu\nu}^{~~I}
\end{eqnarray*}
where, in the last equation we have defined torsion as $T_{\mu\nu}^{~~I}$. Using these, the SO(3,2) Pontryagin density can be written as: 
\begin{eqnarray}
\epsilon^{\mu\nu\alpha\beta}F_{\mu\nu}^{~~AB}(W)F_{\alpha\beta AB}(W)~&=&~-\frac{8}{l^2}\epsilon^{\mu\nu\alpha\beta}\left[(D_{\mu}(\omega) e_\nu^I) ~(D_\alpha(\omega) e_{\beta I})~-~\frac{1}{2}e_\mu^I e_\nu^J R_{\alpha\beta IJ}(\omega) \right]\nonumber\\
&&~+~ \epsilon^{\mu\nu\alpha\beta}R_{\mu\nu}^{~~IJ}(\omega)R_{\alpha\beta IJ}(\omega)
\end{eqnarray}
Evidently, this is the sum of Nieh-Yan and SO(3,1) Pontryagin densities. This identity can be used to express the Lagrangian density (\ref{L-final}) as:
\begin{eqnarray}\label{L-final1}
L(e,\omega)&~=&~\frac{1}{8\kappa} \epsilon^{\mu\nu\alpha\beta}\epsilon_{IJKL}
\left(e_\mu^Ie_\nu^J R_{\alpha\beta}^{~~KL}(\omega)~+~\frac{1}{l^2}e_\mu^Ie_\nu^J e_\alpha^K e_\beta^L\right)\nonumber\\
&~+&~\frac{l^2}{32\kappa}
\epsilon^{\mu\nu\alpha\beta}\epsilon_{IJKL}R_{\mu\nu}^{~~IJ}(\omega)R_{\alpha\beta}^{~~KL}(\omega)
~-~\frac{\eta l^2}{8} \epsilon^{\mu\nu\alpha\beta}F_{\mu\nu}^{~~AB}(W)F_{\alpha\beta AB}(W)
\end{eqnarray}
Thus, the Barbero-Immirzi parameter, which was introduced as a topological coefficient through the Nieh-Yan density in the Lagrangian (\ref{L-ads}), manifests its topological origin through the SO(3,2) Pontryagin density in the final expression above.
%Thus, the theory will depend on this parameter for manifolds (with ALADS boundaries) which have %non-trivial SO(4,1) Pontryagin number. 
This implies that $\eta$ would be present in the corresponding action only for manifolds having a non-zero SO(3,2) Pontryagin index. A manifold of such type has to fall into one of the three classes as given below:

(a) Nieh-Yan number of the manifold is non-zero, but SO(3,1) Pontryagin number is zero;

(b) SO(3,1) Pontryagin number is non-zero, but Nieh-Yan number is zero;

(c) Both Nieh-Yan and SO(3,1) Pontryagin numbers are non-zero.\\
%Evidently, the inclusion of the Nieh-Yan density in the bulk action brings out a far richer structure as %compared to the earlier action formulations without it (see \cite{zanelli}).
These are non-trivial restrictions on the global topology of the manifold.
This is one of the main consequences of our proposed action formulation for ALADS geometries.
\section{Concluding remarks}
We have demonstrated that the inclusion of the Nieh-Yan topological class in the gravity Lagrangian for spacetimes with boundaries leads to a well-defined action formulation. The existence of an extremum of the action is ensured by the addition of appropriate surface terms. Our analysis is sufficiently general in the sense that it applies to spacetimes which can have additional boundaries other than the asymptotic ones. 

In this framework, the topological origin of the Barbero-Immirzi parameter remains manifest throughout, and the inclusion of any arbitrary matter-coupling does not need any additional modification in the bulk action, unlike the earlier formulations based on the Holst action. We also demonstrate that for pure gravity, the boundary contribution from the Nieh-Yan density can be identified exactly with that corresponding to the Holst term. These boundary terms are gauge invariant for both Dirichlet and ALADS boundaries. Thus, the addition of the Nieh-Yan density to the Hilbert-Palatini Lagrangian (along with the corresponding boundary terms) ensures a gauge-invariant Lagrangian as well as a well-defined variational principle. To emphasize, one does not need to introduce the Holst term at any stage of the analysis.

For asymptotic boundaries which are locally AdS (or dS), the Lagrangian with the Nieh-Yan density admits an extremum if the other two topological densities, i.e. Euler and Pontryagin, are also included with fixed coefficients. Thus, although the full Lagrangian density contains all three topological densities which exist in four dimensional gravity theory\cite{date,kaul}, it has only one independent topological parameter, namely, the Barbero-Immirzi parameter $\eta$. In the final analysis, it emerges as a coefficient of the SO(3,2) (or SO(4,1)) Pontryagin topological density in the Lagrangian. Thus, $\eta$ would be relevant in the action principle only for those ALADS manifolds which have a nontrivial SO(3,2) Pontryagin index. This fact also provides a potentially interesting hint as to how the quantum theory corresponding to gravity with a cosmological constant might perceive this topological parameter.

\acknowledgments
The author is indebted to Romesh Kaul for his comments, and to Miguel Campiglia for his critical reading of the manuscript. Discussions with Amit Ghosh, Kumar Gupta and Joseph Samuel are gratefully acknowledged.

%\begin{eqnarray}
%\end{eqnarray}
%

\end{document}